# Quantized charge fractionalization at quantum Hall Y junctions in the disorder-dominated regime

Chaojing Lin[1,2,*], Masayuki Hashisaka[3], Takafumi Akiho[3], Koji Muraki[3], and Toshimasa Fujisawa[1,*]

[1]*Department of Physics, Tokyo Institute of Technology, 2-12-1 Ookayama, Meguro, Tokyo 152-8551, Japan.*
[2]*Tokyo Tech Academy for Super Smart Society, Tokyo Institute of Technology, 2-12-1 Ookayama, Meguro, Tokyo 152-8551, Japan.*
[3]*NTT Basic Research Laboratories, NTT Corporation, 3-1 Morinosato-Wakamiya, Atsugi, Kanagawa 243-0198, Japan.*

**Fractionalization is a phenomenon where an elementary excitation partitions into several pieces. This picture explains non-trivial transport through a junction of one-dimensional edge channels defined by topologically distinct quantum Hall states, for example, a hole-conjugate state at Landau-level filling factor $\nu$ = 2/3. Here we employ a time-resolved scheme to identify an elementary fractionalization process; injection of charge $q$ from a non-interaction region into an interacting and scattering region of one-dimensional channels results in the formation of a collective excitation with charge $(1-r)q$ by reflecting fractionalized charge $rq$. The fractionalization factors, $r$ = 0.34±0.03 for $\nu$ = 2/3 and $r$ = 0.49±0.03 for $\nu$ = 2, are consistent with the quantized values of 1/3 and 1/2, respectively, which are expected in the disorder dominated regime. The scheme can be used for generating and transporting fractionalized charges with a well-defined time course along a well-defined path.**

One-dimensional electronic systems provide non-trivial many-body effects that cannot be explained with single-particle pictures[1]. Theoretically, these effects can be calculated using bosonization techniques and the bosonic (plasmonic) scattering approach, which have been applied for both dc and ac responses even in inhomogeneous and disordered systems[1-6]. Experimentally, many-body states can be investigated using electronic and optical techniques[7-10]. Among them, one-dimensional edge channels in integer and fractional quantum Hall (QH) systems[11-14] are attractive for studying non-trivial excitations in multiple channels by utilizing mesoscopic devices[15-18]. The focus of this study is transport eigenmodes that govern the interacting edge channels.

For example, the charge and spin (dipolar) modes for copropagating channels in the integer QH system at $\nu$ = 2 were investigated based on the Coulomb interaction in terms of the chiral Tomonaga-Luttinger liquid[19-21]. At a Y-junction where two decoupled channels join to form an interacting region, an electronic excitation incident from the non-interacting region is fractionalized into non-trivial charge and spin excitations in the interacting region[19,22-24]. In the absence of interchannel tunneling, the eigenmodes are determined by the interaction parameters and can hence deviate from the pure charge and spin modes. In this interaction-dominated regime, the fractionalization ratio assumes a non-universal interaction-dependent value, as demonstrated in frequency- and time-resolved measurements as well as noise measurements[25-27].

A similar class of coupled modes appears when disorder allows for significant tunneling between two edge channels. A well-known example is the charge and neutral modes in the 'hole conjugate' fractional QH state at $\nu$ = 2/3, as suggested by noise measurements and transport properties for short interacting regions[28-33]. We assumed a reconstructed edge with counterpropagating integer and fractional channels[12,13], whereas alternative effective models can be considered[34,35]. Theoretically, the charge and neutral modes appear at the Kane-Fisher-Polchinski fixed point in the renormalization group flow[36]. In this disorder-dominated regime, an elementary excitation should be fractionalized into pure charge and neutral modes with a quantized ratio at a Y junction of interacting and non-interacting regions[36,37].

In this study, we have experimentally identified this quantized fractionalization ratio by employing time-resolved measurements for the hole-conjugate fractional state at $\nu$ = 2/3. A similar quantized fractionalization is also found in the integer QH state at $\nu$ = 2 in the presence of significant tunneling. The obtained feature is supported by a simulation involving a realistic model based on the plasmon scattering approach. The quantized charge fractionalization describes the dc characteristics as well.

## Results

**Fractionalization processes.** We first consider the edge of the fractional state at $\nu$ = 2/3, where the counterpropagating $\Delta\nu$ = 1 and 1/3 one-dimensional channels[12] are formed along the interface to the electronic vacuum ($\nu$ = 0), as shown in Fig. 1a. Here, $\Delta\nu = |\nu_1 - \nu_2|$ denotes a channel along an interface between insulating (incompressible) regions with $\nu = \nu_1$ and $\nu_2$. Disorder-induced scattering renders them describable as a composite $\Delta\nu$ = 2/3 channel with two counterpropagating transport modes[36], i.e. a charge mode carrying a charge and a neutral mode carrying heat. We address fractionalization processes at Y-junctions comprising $\Delta\nu$ = 1, 2/3, and 1/3 channels, as shown in Figs. 1b-d. Two types of Y-junctions are possible, i.e. $Y_C$ and $Y_N$, which form depending on the cyclic order of the insulating regions and the direction of the magnetic field $B$. For the configuration shown in Fig. 1b, a wave packet of charge $q$ incident from the $\Delta\nu$ = 1 channel is fractionalized with factor $r$ (= 1/3 in the disorder dominated regime) at junction $Y_C$ into fractional charges $(1-r)q$ and $rq$, which propagate through the $\Delta\nu$ = 2/3 and $\Delta\nu$ = 1/3 channels, respectively. This occurs because the charge mode in the $\Delta\nu$ = 2/3 channel is composed of charges $q$ in the $\Delta\nu$ = 1 channel and $-rq$ in the $\Delta\nu$ = 1/3 channel[36]. The formation of this collective excitation requires a charge $rq$ to be reflected back into the uncoupled $\Delta\nu$ = 1/3 channel[37]. A similar reflection is expected when a wave packet of charge $q$ is injected from a $\Delta\nu$ = 1/3 channel to junction $Y_N$ shown in Fig. 1c, where neutral excitation in the $\Delta\nu$ = 2/3 channel is formed by reflecting charge $q$ in the downstream $\Delta\nu$ = 1 channel. As shown in Fig. 1d, a charge wave packet in the charge mode of the $\Delta\nu$ = 2/3 channel is decomposed into a charge in the $\Delta\nu$ = 1 channel and heat in the neutral mode. We focus on the charge fractionalization by neglecting neutral excitations as the length of the $\Delta\nu$ = 2/3 channel ($L$ > 100 μm) is much longer than the equilibration length $l_{eq}$ (typically ~ 10 μm)[32,33].

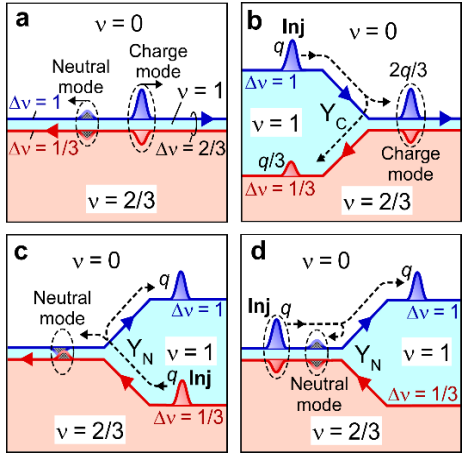

**Figure 1. QH Y-junctions formed with $\Delta\nu = 1$, 2/3, and 1/3 channels. a**, The charge and neutral modes in the disorder dominated regime of a composite $\Delta\nu = 2/3$ channel comprising $\Delta\nu = 1$ (blue) and 1/3 (red) channels along the boundaries of a hole-conjugate QH region with $\nu = 2/3$, a narrow integer state with $\nu = 1$, and the electronic vacuum ($\nu = 0$). Excitations are represented by positive and negative wave packets with ratios of charges in $\Delta\nu = 1$ and 1/3 channels (1:-1/3 for the charge mode and 1:-1 for the neutral mode). **b**, Charge fractionalization at junction $Y_C$. An incoming wave packet with charge $q$ in the $\Delta\nu = 1$ channel is fractionalized into two packets with $2q/3$ (comprising $q$ and $-q/3$) in the $\Delta\nu = 2/3$ channel and $q/3$ in the $\Delta\nu = 1/3$ channel. **c-d**, Neutral reflections at junction $Y_N$. An injected packet with charge $q$ in the $\Delta\nu = 1/3$ channel splits into charge $q$ in the $\Delta\nu = 1$ channel and neutral excitation (comprising $-q$ and $q$) in the $\Delta\nu = 2/3$ channel in **c**, and so as the packet in the $\Delta\nu = 2/3$ channel in **d**.

**Quantized fractionalization in $\nu = 2/3$ case.** We demonstrate the charge fractionalization in time-domain measurements using several devices formed in a standard AlGaAs/GaAs heterostructure (see Methods and Supplementary Note 1). The following data were obtained at ~50 mK from devices #1 and #2 fabricated on the same chip, as schematically shown in Fig. 2a. For device #1, two Y-junctions, $Y_C$ and $Y_N$, formed at the intersections of the three regions—the ungated region with bulk filling factor $\nu_B = 2/3$, the gated region with a tunable $\nu_G$ (= 1 in Fig. 2a), and vacuum. An initial charge wave packet was excited by applying a voltage step to the injector gate $G_I$, and the waveforms of the charge packets after passing through the junctions were investigated by applying a voltage pulse of width $t_w$ (0.08 – 0.15 ns) to the detector gate $G_D$. Charge waveforms were obtained by measuring the detector current $I_D$ at various time delays $t_d$ of the voltage pulse with respect to the voltage step (see Methods)[38]. Trace (i) in Fig. 2b is a reference showing that a single charge packet was observed for $\nu_G = 0$ (the gate voltage $V_g = -0.3$ V), i.e. when a single $\Delta\nu = 2/3$ channel without Y-junctions is formed, as shown in the inset. This is a typical characteristic of the edge magnetoplasmon mode[39-41] at $\nu = 2/3$. When the $Y_C$ and $Y_N$ junctions were activated by setting $\nu_G = 1$ ($V_g = +0.26$ V, $B = 11.5$ T), a clear charge fractionalization manifested as two distinct packets in trace (ii). The first packet is associated with the direct propagation through junction $Y_N$, $\Delta\nu = 1$ channel, and junction $Y_C$. The second one is delayed by the round trip around the gated region, as illustrated in the insets. Subsequent packets associated with further fractionalization processes are extremely small to be resolved. By assuming $r = 1/3$, the entire process yields a series of packets with $2q/3$, $2q/9$, … toward the detector. We evaluated the charge $q_t$ in the reference wave packet in (i) as well as $q_{f1}$ and $q_{f2}$ in the first and second packets in (ii), respectively, from the area of the peaks. The obtained $q_t$, $q_{f1}$, and $q_{f2}$ are plotted in Fig. 3b as a function of $V_g$, with the vertical axis normalized by the $q_t$ value at $V_g = -0.3$ V ($\nu_G = 0$). The ratios $q_{f1}/q_t$ and $q_{f2}/q_t$ are similar to the expected values of $(1 - r) = 2/3$ and $(1 - r)r = 2/9$, respectively, when the $\Delta\nu = 1$ and $1/3$ channels are well defined at $\nu_G \geq 1$. In particular, $r = q_{f2}/q_{f1}$ estimated from each $I_D$ profile yields $r = 0.34\pm0.03$ in the range of $V_g = 0.21 - 0.27$ V, as shown in the inset of Fig. 3b, consistent with the quantized value of 1/3.

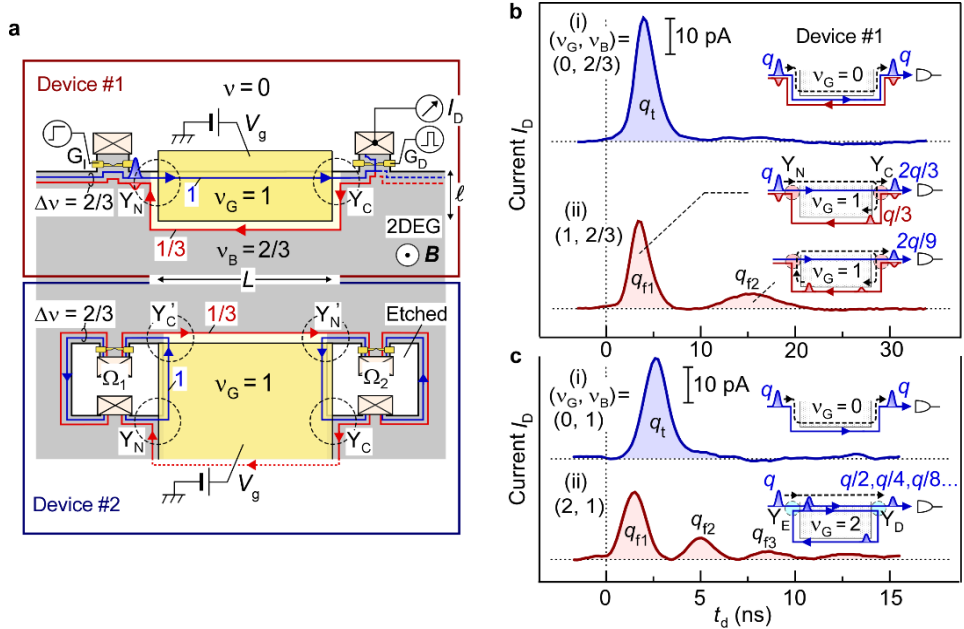

**Figure 2. Quantized fractionalization of charge wave packets. a**, Measurement setup with devices #1 and #2. Application of voltage $V_g$ to the large gate (yellow) forms a rectangular QH region ($L = 300$ μm and $\ell = 20$ μm for #1) and QH junctions $Y_N$ and $Y_C$ at $\nu_G = 1$ and $\nu_B = 2/3$. For #1, an initial wave packet is excited by applying a voltage step to the injector gate $G_I$ with the underneath fully depleted. Fractionalized wave packets are detected by applying a voltage pulse of width $t_w$ to transmit a part of the packet to be detected as current $I_D$. For #2, two-terminal dc conductance $G$ through similar Y-junctions is measured with ohmic contacts in Corbino geometry. **b**, Typical charge waveforms obtained in current $I_D$ as a function of delay time $t_d$ of the detector voltage pulse with respect to the excitation voltage step. The reference trace (i) at $\nu_G = 0$ ($V_g = -0.3$ V) and trace (ii) showing charge fractionalizations at $\nu_G = 1$ ($V_g = 0.26$ V) were obtained at $\nu_B = 2/3$ ($B = 11.5$ T). Areas under the peaks represent charges with ratios $q_{f1}/q_t \sim 2/3$ and $q_{f2}/q_t \sim 2/9$. **c**, The reference trace (i) at $\nu_G = 0$ ($V_g = -0.3$ V) and trace (ii) showing fractionalizations at $\nu_G = 2$ ($V_g = 0.34$ V) obtained at $\nu_B = 1$ ($B = 7.5$ T). The areas under the peaks show $q_{f2}/q_{f1} \sim 1/2$ and $q_{f3}/q_t \sim 1/4$. Propagation of charge wave packets is illustrated in the respective insets.

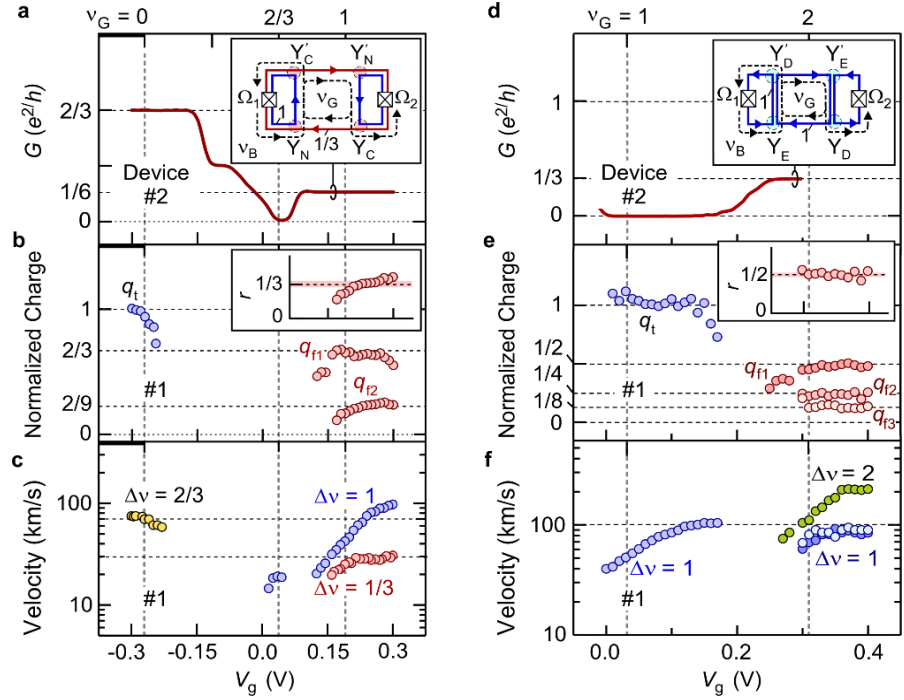

**Figure 3**. **Characteristics of charge transport. a** and **d** $V_g$-dependence of two-terminal conductance $G$ measured with ohmic contacts $\Omega_1$ and $\Omega_2$ of device #2 obtained at $\nu_B = 2/3$ in **a** and $\nu_B = 1$ in **d**. The insets show the channel configurations, where multiple charge fractionalizations at Y-junctions explain the plateau $G = e^2/6h$ at $\nu_G = 1$ in **a** and $G = e^2/3h$ at $\nu_G = 2$ in **d**. **b** and **e** The reference charge $q_t$, and fractionalized charges $q_{f1}$, $q_{f2}$, and $q_{f3}$ in the respective packets normalized by $q_t$. A single reference packet typically involves $q_t \cong 240e$ in **b** and $30e$ in **e**. The clear plateaus of $q_f/q_t$ indicate the quantized fractionalization. The insets show fractionalization factor $r = q_{f2}/q_{f1}$ with a constant region ($r = 0.34 \pm 0.03$ in **b** and $r = 0.49 \pm 0.03$ in **e**). **c** and **f** Charge velocities of the channels. The $\Delta\nu = 1/3$ interface channel between $\nu = 1$ and $2/3$ regions, the $\Delta\nu = 1$ edge channel between $\nu_G = 1$ and vacuum, and the $\Delta\nu = 2/3$ composite channel between $\nu_G = 0$ and $\nu_B = 2/3$ are shown in **c**, whereas the $\Delta\nu = 1$ interface channel between $\nu_G = 2$ and $\nu_B = 1$ regions (0.3 V < $V_g$ < 0.4 V, estimated from the second and the third packets), the $\Delta\nu = 1$ edge channel between $\nu_G = 1$ and vacuum ($V_g < 0.18$ V), and the $\Delta\nu = 2$ composite channel between $\nu_G = 2$ and vacuum are shown in **f**. Data in **b**, **c**, **e**, and **f** were obtained using device #1. Vertical dotted lines for representative $\nu_G$ values were determined from a separate four-terminal measurement (see Supplementary Note 1).

This observation is supported by the dc characteristics of device #2, which has Corbino geometry with ohmic contacts surrounded by a QH state, as shown in the lower part of Fig. 2a. Transport through the $\Delta\nu = 1/3$ channel formed between $\nu_G = 1$ and $\nu_B = 2/3$ regions involves the equilibration associated with scattering between the coupled $\Delta\nu = 1$ and $1/3$ channels inside the composite $\Delta\nu = 2/3$ channels. Figure 3a shows the two-terminal conductance $G$ between ohmic contacts $\Omega_1$ and $\Omega_2$ with other ohmic contacts floating. The clear plateau of $G \cong e^2/6h$ at $V_g \sim$ +0.2 V ($\nu_G = 1$) ensures a full equilibration in the $\Delta\nu = 1/3$ channel and negligible backscattering in both $\nu_G = 1$ and $\nu_B = 2/3$ regions. This is a requisite for clear quantization of $r = 1/3$. Whereas the dc characteristics of systems involving composite $\Delta\nu = 2/3$ channels have been successfully explained in various ways[32,33,37], we herein demonstrate that the same can also be understood with the quantized charge fractionalization. As shown by the simplified channel configuration in the inset of Fig. 3a, a fictitious charge packet $q$ emanating from $\Omega_1$ is fractionalized into a series of charge packets through the paths shown by the dashed lines. Some of them reach $\Omega_2$ with the first charge $2q/9$ through path $\Omega_1$ - $Y_N$ - $Y_C$' - $Y_N$' - $Y_C$ - $\Omega_2$, followed by others multiplied by the geometric ratio of 1/9 associated with round trip $Y_C$ - $Y_N$ - $Y_C$' - $Y_N$' - $Y_C$. The total charge $q/4$ reaching $\Omega_2$ explains $G = e^2/6h$ for the conductance $2e^2/3h$ of the source channels connected to $\Omega_1$ and $\Omega_2$. Hence, charge fractionalization provides a unified view of dc and time-dependent charge transport.

**Quantized fractionalization in $\nu = 2$ case.** We observed similar quantized fractionalization with integer QH states at $\nu_G = 2$ and $\nu_B = 1$, when the two $\Delta\nu = 1$ channels with up- and down-spins were prepared in the disorder-dominated regime. The two channels are coupled to form a composite $\Delta\nu = 2$ channel, as shown in the bottom inset of Fig. 2c. Significant scattering between them is allowed for example by coupling to nuclear spins[42]. Separate experiments show full equilibration for a channel length of ~300 μm in device #2 (see Supplementary Note 2). Our previous study showed a short equilibration length of ~10 μm in a similar device with a slightly lower electron density[33]. In this disorder-dominated regime, the transport eigenmodes of the $\Delta\nu = 2$ channel should be a pure symmetric charge mode and a short-lived antisymmetric neutral mode (see Methods). These modes are excited at junction $Y_E$ and decomposed at junction $Y_D$ with quantized charge fractionalization of factor $r = 1/2$. Namely, a single charge packet with $q$ in the symmetric mode splits into two packets with $(1-r)q$ and $rq$ in the up- and down-spin channels, respectively. Compared with the reference trace (i) in Fig. 2c for $(\nu_G, \nu_B) = (0, 1)$, trace (ii) shows charge fractionalizations for ($\nu_G$, $\nu_B) = (2, 1)$ at $V_g = +0.34$ V and $B = 7.5$ T. A series of well-isolated packets, $q_{f1}, q_{f2}, ...$, manifests the multiple fractionalization processes at $Y_D$. As plotted in Fig. 3e, the fractionalization factor $r = q_{f1}/q_{f2} = 0.49 \pm 0.03$ obtained in the range of $V_g = 0.31 - 0.37$ V is consistent with the quantized value of 1/2. This is in contrast to previous studies pertaining to the interaction dominated regime, where asymmetric modes with an interaction dependent factor of $r \sim 0.4$) were observed[26,27].

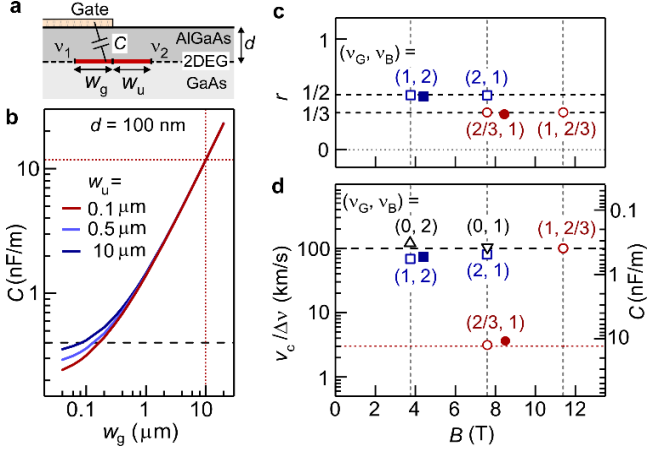
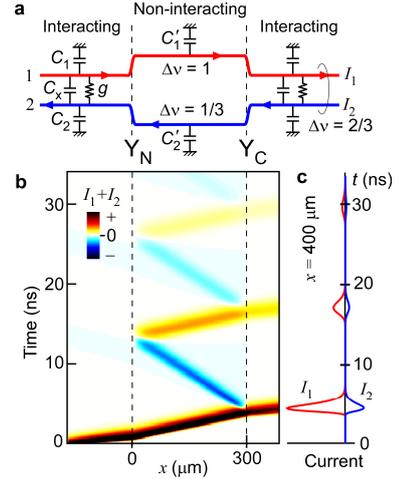

**Figure 4. Velocity of the interface mode. a**, Schematic cross-section around the interface channel $\Delta\nu = |\nu_1 - \nu_2|$ of width $w_g + w_u$ ($w_g$ in the gated region and $w_u$ in the ungated region) between two QH states at $\nu_1$ and $\nu_2$. The interaction inside the channel can be described with geometric capacitance $C$ to the gate. **b**, Calculated capacitance $C$ as a function of $w_g$ for several $w_u$ values. **c**, Fractionalization factor $r$ for junction $Y_C$ obtained at $(\nu_G, \nu_B) = (2/3, 1)$ and $(1, 2/3)$ showing $r \sim 1/3$ (circles), and junction $Y_D$ at $(1, 2)$ and $(2, 1)$ showing $r \sim 1/2$. **d**, Normalized charge velocities $v_c/\Delta\nu$ for fractional $\Delta\nu = 1/3$ interface channels at $(2/3, 1)$ and $(1, 2/3)$ marked with circles, integer $\Delta\nu = 1$ interface channels at $(1, 2)$ and $(2, 1)$ marked with squares, and conventional edge channels at $(0, 2)$ and $(0, 1)$ marked with triangles. Data obtained after light irradiation are marked with solid symbols. Channel capacitance $C$ is shown on the right scale.

We observed a clear two-terminal conductance plateau $G = e^2/3h$ at $(\nu_G, \nu_B) = (2, 1)$ using device #2, as shown in Fig. 3d. This conductance is 1/3 of the original $G = e^2/h$ of the single integer channel emanating from the ohmic contacts. This can be understood as the sum of the first transmission coefficient (the square of the fractionalization factor 1/2) of a fictitious charge packet through path $\Omega_1 - Y_E - Y_D' - Y_E' - Y_D - \Omega_2$ followed by others with a geometric ratio of 1/4 associated with round trip $Y_D - Y_E - Y_D' - Y_E' - Y_D$, as shown in the inset. Hence, the quantized fractionalization also explains the dc characteristics of the integer channels.

**Plasmon velocities.** The velocity of the wave packet is an important parameter that reflects the interaction, as evident from chiral Tomonaga-Luttinger theories[2,10,19]. We experimentally estimated the velocities from the time of flight, as summarized in Figs. 3c and 3f. The velocities of the edge channels ($\Delta\nu = 1$ channel between $\nu_G = 1$ and vacuum and $\Delta\nu = 2/3$ channel between $\nu_G = 0$ and $\nu_B = 2/3$ in Fig. 3c) are comparable to those in previous reports regarding edge magnetoplasmons[38,39,43,44]. The velocity of the $\Delta\nu = 1/3$ interface channel between the $\nu = 1$ and $2/3$ regions, $\sim 30$ km/s, is particularly important for transporting fractional charges[45]. Unlike edge channels with a well-defined confining potential, the interface channel is supported by two QH states with a slight difference in their electrostatic potentials. Therefore, the contribution of the single-particle drift velocity arising from the potential gradient is negligible. This is particularly relevant to the $\Delta\nu = 1/3$ channel, as the Fermi level remains in the lowest Landau level in the fractional state.

**Figure 5. Charge fractionalization calculated using a plasmon model a**, Non-interacting edge channels $\Delta\nu = 1$ and $\Delta\nu = 1/3$ in the central region and composite $\Delta\nu = 2/3$ channels in the interaction regions on both sides, forming junctions $Y_N$ and $Y_C$. **b**, Time evolution of a charge wave packet initially prepared in the left interacting region at $x = -200$ μm, showing full transmission and full reflection at $Y_N$ ($x = 0$) and charge fractionalization at $Y_C$ ($x = 300$ μm). The sum of the currents $I_1$ and $I_2$ in the original $\Delta\nu = 1$ and 1/3 channels, respectively, is plotted in a color scale. The numerical simulation was performed using realistic parameters: capacitances $C_1 = C_2 = C_x = 0.07$ nF/m; scattering conductance $g$ inducing $l_{eq} = 10$ μm in the interacting regions; $C_1' = C_2' = 0.4$ nF/m in the non-interaction region. **c**, Time-dependent $I_1$ and $I_2$ at $x = 400$ μm in right interaction region. Each packet shows $I_2 = -I_1/3$ of the charge mode.

To understand the origin of the velocity, we assume that the charge velocity of a $\Delta\nu$ channel is expressed as $v_c = \Delta\nu g_q/C$, where $g_q = e^2/h$ is the quantized conductance, and $1/C$ measures the interaction[16]. Practically, $C$ should be dominated by the geometric capacitance (per unit length) between the channel and a nearby gate[46]. For an interface channel along the side of the gate shown in Fig. 4a, this $C$ is expected to depend on the width, $w = w_g + w_u$, of the channel (compressible region), where $w_g$ ($w_u$) is the spread under the gate (in the ungated region). Our numerical simulation (see Methods) shows that $C$ is determined primarily by $w_g$ rather than $w_u$ (Fig. 4b). The normalized velocities, $v_c/\Delta\nu$, obtained for various values of $(\nu_G, \nu_B)$, are summarized in Fig. 4d. Here, the data for $\nu_G > \nu_B$ and $\nu_G < \nu_B$ were obtained using devices #1 and #2, respectively, with $V_g > 0$ and $V_g < 0$ (see Supplementary Notes 2 and 3). Except for $(\nu_G, \nu_B) = (2/3, 1)$, $v_c/\Delta\nu$ indicates similar values for all interface channels, i.e. $\Delta\nu = 1/3$ (circles) and 1 (squares), as well as edge channel $\Delta\nu = 1$ (triangles). This coincidence suggests that the velocities are determined by a similar $C \sim 0.4$ nF/m, as shown on the right axis. A comparison with Fig. 4b implies that $w_g$ is sufficiently narrow, comparable to the depth $d \sim 100$ nm of the electron system from the surface. This indicates that the velocity $\sim 30$ km/s of the $\Delta\nu = 1/3$ channel obtained for $(\nu_G, \nu_B) = (1, 2/3)$ is reasonable. Meanwhile, a significantly lower velocity of $\sim 1$ km/s was observed for the $\Delta\nu = 1/3$ channels in the $(2/3, 1)$ configuration. This suggests a wide $w_g \sim 10$ μm in the crude model or quasi-diffusive transport in the presence of disorder potential. Whereas this might be related to the small energy gaps of the QH states at lower $B$ in this configuration, the velocity did not increase significantly with $B$ even after light irradiation (solid circle), which increased the electron density. The former (1, 2/3) configuration with a fractional state in an ungated region might be suitable for minimizing the time-of-flight and hence decoherence in a fractional-charge interferometer. It is noteworthy that the fractionalization factor $r$ summarized in Fig. 4c remained at approximately 1/3 even when the velocity reduced significantly.

**Discussion**

The observation above suggests robust fractionalization factors in the disorder-dominated regime. This is consistent with the plasmon (charge density wave) transport model (see Methods) shown in Fig. 5a, where interaction and scattering are characterized by distributed capacitances and scattering conductances, respectively[46-48]. The transport eigenmodes generally deviate from the pure charge and neutral modes at higher frequencies. However, the deviation is small in the low-frequency regime, where the wavelength $\lambda$ of the plasmon is much greater than the equilibration length $l_{eq}$. This is observed in the numerical simulation of multiple charge fractionalizations with realistic parameters, as shown in Figs. 5b and 5c, where the distortion of the charge waveform is negligible. The obtained narrow width (a few nanoseconds) of the fractionalized wave packets encourages studying microscopic fractionalization processes including neutral modes and heat generation, which can be used to identify the appropriate effective model[34-36,49]. The deterministic fractionalization processes may benefit the search for non-trivial anyonic statistics of fractional charges[48,50-53].

**Methods**

**Device fabrication.** The devices were fabricated from a standard GaAs/AlGaAs heterostructure with a two-dimensional electron gas (2DEG) located 100 nm below the surface having an electron density of $1.85\times10^{11}$ cm$^{-2}$ in the dark and $2.07\times10^{11}$ cm$^{-2}$ after light irradiation at low temperature. After patterning holes into the 2DEG for the Corbino geometry, ohmic contacts were formed by alloying Au–Ge–Ni metal films; subsequently metal gates were patterned using photolithography and electron-beam lithography (see Supplementary Note 1 for details).

**Time-of-flight experiment.** A charge wave packet was generated by depleting electrons near the injection gate $G_I$ of length $l_I \sim 50$ μm by applying a voltage step $\Delta V_I = 5 - 15$ mV to the static voltage of $-0.2 – -0.3$ V. This induced charge $q_I \sim C_I l_I \Delta V_I$ in the packet, where $C_I l_I$ is the coupling capacitance. The charge waveform $\rho(t)$ was evaluated by applying a detector pulse $\Delta V_D = 20$ mV to the static voltage of $-0.3 – -0.4$ V on gate $G_D$ to change the transmission probability to the detector ohmic contact by $\Delta T_D \sim 0.17$. This induced a detector current $I_D = \Delta T_D \rho(t) t_w/T_{rep}$ with repetition time $T_{rep}$ of the voltage step and the pulse of a width $t_w = 0.08 - 0.15$ ns. The charge in the wave packet was estimated from the integrated current. The time origin of the delay $t_d$ was calibrated from a similar experiment at zero magnetic field, where the wave packet propagates much faster with a velocity on the order of $10^7$ m/s[16,26,38].

The charge velocity was estimated from the time-of-flight. For the wave packets shown in Fig. 2b, velocities $v_{2/3,u}$ and $v_{1,u}$ of the $\Delta \nu = 2/3$ and 1 channels under the gate (length $L$) were estimated from the time-of-flight of the first wave packet in traces (i) and (ii), respectively, by disregarding the short time-of-flight ($\sim 0.5$ ns) in the $\Delta \nu = 2/3$ ungated channel[39]. Subsequently, the velocity $v_{1/3,s}$ of the $\Delta \nu = 1/3$ channel along the side of the gate (length $L + 2\ell$) was estimated from the delay of the second wave packet in trace (ii) and the predetermined $v_{1,u}$. Because the velocity depends strongly on the electrostatic environment, the channels formed along the side of the gates were compared, as shown in Fig. 4d.

**Capacitance of interface channel.** The interface channel is a compressible stripe of finite width $w$ between two incompressible regions. As the electrostatic potential for this situation is challenging[54], we assumed finite widths $w_g$ and $w_u$ in the gated and ungated regions, respectively, as shown in Fig. 4a. By considering the incompressible regions as insulators, the capacitance between the channel and the gate was calculated using commercial software COMSOL based on the finite-element method.

**Fractionalization factor at high frequencies** We used the plasmon scattering approach to simulate the fractionalization process in the presence of disorder-induced tunneling[4,46,47]. Consider two one-dimensional chiral channels ($n = 1$ and 2) with conductance $\sigma_n$ (positive for right movers and negative for left movers), as shown in Fig. 5a. The charge density $\rho_n$, electrochemical potential $V_n$, and current $I_n = \sigma_n V_n$ are related to each other with the Coulomb interaction characterized by the self-capacitance $C_n$ (to the ground) and coupling capacitance $C_X$ per unit length[16,26]. The quantum capacitance was absorbed in those capacitances. Scattering between the channels was considered with scattering conductance $g$ per unit length[47]. Based on current conservation, we derived the following wave equation:

$$-\frac{\partial}{\partial x}\begin{pmatrix}I_1\\I_2\end{pmatrix} = \left[\begin{pmatrix}C_1 + C_x & -C_x\\-C_x & C_2 + C_x\end{pmatrix}\frac{\partial}{\partial t} + g\begin{pmatrix}1 & -1\\-1 & 1\end{pmatrix}\right]\begin{pmatrix}V_1\\V_2\end{pmatrix} \quad (1)$$

Transport eigenmodes $\hat{I}_m = (\tilde{I}_1, \tilde{I}_2)^\mathsf{T}$ can be calculated for alternating current $I_n = \tilde{I}_n e^{i(kx-\omega t)}$ with amplitude $\tilde{I}_n$ at frequency $\omega$. The resulting $k$ (complex) for each mode measures the wavenumber in the real part and the decay rate in the imaginary part. For the fractional case with $\sigma_1 = e^2/h$, $\sigma_2 = -e^2/3h$, and $g > 0$, pure charge and neutral modes with $\tilde{I}_2/\tilde{I}_1 = -1/3$ and $-1$, respectively, appeared at $g >> \omega(C_1 + 3C_2)$ in the disorder-dominated regime, and interaction-dependent modes appeared at $g << \omega(C_1 + 3C_2)$ in the interaction-dominated regime. Because the solution in the zero-frequency limit ($\omega \to 0$) provides the equilibration length $l_{eq} = \sigma_q/2g$, the disorder-dominated regime corresponds to the plasmon wavelength $\lambda$ much longer than $l_{eq}$. Our wave packet contains a long wavelength in the Fourier components ($\lambda \gtrsim 800$ μm in the $\Delta \nu = 2/3$ channel for the data in Fig. 2b and $\lambda \gtrsim 300$ μm in the $\Delta \nu = 2$ channel for the data in Fig. 2c). Hence, all data shown herein are obtained from the disorder-dominated regime for our sample with $l_{eq} \sim 10$ μm. In this case, the charge mode exhibits a slight decay with an angle $\arg[k] \sim 2\pi (l_{eq}/\lambda)(C_1 + 3C_2)^2/(C_1 + C_2)^2$ in the lowest order. This broadens the wave packet only slightly. The time evolutions of $I_1$ and $I_2$ in Fig. 5 were obtained by integrating Eq. (1) with current conservation at the boundaries of non-interacting ($C_X = 0$ and $g = 0$) and interacting regions.

**Supplementary Notes**

**Supplementary Note 1: Sample and low-frequency transport characteristics**

Supplementary Fig. 1a shows an optical micrograph of the chip that we investigated. The upper and lower parts of the chip were used as devices #1 and #2, respectively. The AlGaAs/GaAs heterostructure with a two-dimensional electron gas (2DEG) was partially removed by chemical etching (dark regions). Whereas several ohmic contacts $\Omega_I, \Omega_{II}, \ldots$ were formed on the outer edge of the 2DEG, a few ohmic contacts $\Omega_1, \Omega_1', \Omega_2, \Omega_2', \ldots$ were formed on the inner edges to provide the Corbino geometry. The bulk filling factor $\nu_B$ in the ungated region was set by applying a perpendicular magnetic field $B$. By applying gate voltage $V_g$ to large gates $G_1$ and $G_2$, the filling factor $\nu_G$ under the gated regions was selected. Small gates $G_I$, $G_I'$, $G_D$, and $G_D'$ were used as charge injectors and detectors for devices #1 and #2.

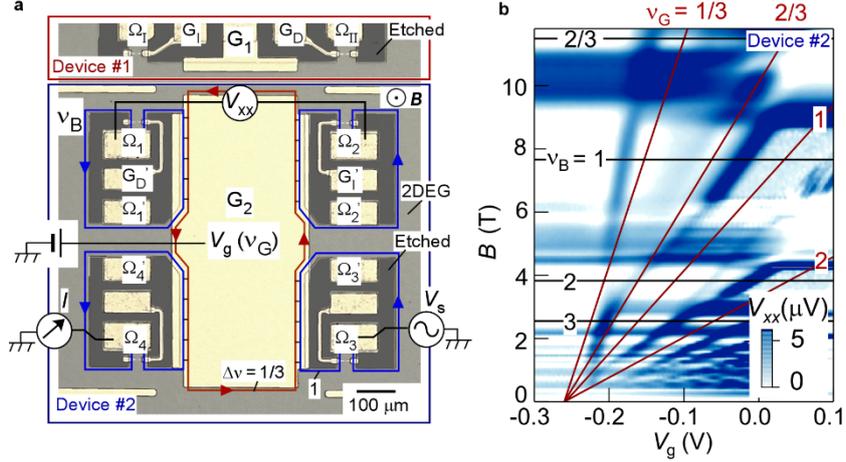

**Supplementary Figure 1. Low-frequency characteristics of the QH states. a**, Optical micrograph of devices #1 and #2 with measurement setup for four-terminal dc measurements. **b**, Color plot of $V_{xx}$ as a function of gate voltage $V_g$ and magnetic field $B$. The QH states are labeled $\nu_B$ for the bulk (horizontal lines) and $\nu_G$ under the gate (inclined lines).

Four-terminal dc conductance measurements were performed using the setup shown in Supplementary Fig. 1a. With source voltage $V_s = 30$ μV (37 Hz) applied to ohmic contact $\Omega_3$, the two-terminal conductance $G$ (= $I/V_s$) was obtained by measuring the current $I$ at $\Omega_4$. In addition, the longitudinal voltage $V_{xx}$ was measured using $\Omega_1$ and $\Omega_2$. Supplementary Fig. 1b shows a color plot of the measured $V_{xx}$ as a function of gate voltage $V_g$ and magnetic field $B$ under dark conditions. The overall patterns in $V_{xx}$ can be understood with the variation in $\nu_B$ in the bulk shown by horizontal lines and $\nu_G$ under the gate shown by inclined lines. Vanishing $V_{xx}$ regions (white regions) show negligible bulk scattering in both the gated and ungated QH states. We assumed the same QH states were formed under gate $G_1$ in device #1.

For example, a system with $\nu_B = 1$ and $\nu_G = 2/3$ was prepared at $V_g = -0.08$ V and $B = 7.5$ T. In this case, a complex $\Delta\nu = 2/3$ channel made of counterpropagating $\Delta\nu = 1/3$ and 1 channels was formed by edge reconstruction with a non-monotonic variation of $\nu$ from 0 through 1 to 2/3[12]. Consequently, the single $\Delta\nu = 1/3$ channel yielded a closed loop along the side of gate $G_2$, as shown by the red line in Supplementary Fig. 1a. This channel is coupled to four $\Delta\nu = 1$ channels (blue) connected to ohmic contacts via the complex $\Delta\nu = 2/3$ channels (parallel blue and red lines). Transport is allowed by scattering in the $\Delta\nu = 2/3$ channels. We find that the two-terminal conductance $G \cong 1/3 \, e^2/h$ (not shown) agrees well with the full equilibration. The length of the complex $\Delta\nu = 2/3$ channel, ~ 300 μm, is longer than typical equilibration length, ~ 10 μm obtained in our previous study for a similar 2DEG wafer[33].

**Supplementary Note 2: Charge waveforms obtained from device #1**

Supplementary Fig. 2a shows the current $I_D$ as a function of delay time $t_d$ in a wide range of $V_g$ from -0.3 V to +0.3 V, obtained at $\nu_B = 2/3$ ($B = 11.5$ T) in device #1. A single wave packet was observed at $V_g \sim -0.27$ V ($\nu_G = 0$), where the packet propagated along the composite $\Delta\nu = 2/3$ channel comprising $\Delta\nu = 1$ (blue) and 1/3 (red) channels along the perimeter of the gate, and $V_g = +0.03$ V ($\nu_G = 2/3$), where the packets propagated along the etching step under the gate, as shown in the insets. The former indicates a faster velocity of ~ 70 km/s (short time-of-flight for a long distance) than the latter (~ 20 km/s). This can be understood based on the screening effect of the metal gate[39,41].

When $V_g$ was increased above +0.18 V ($\nu_G = 1$), two distinct packets appeared in the $I_D$ profile. They are associated with multiple fractionalizations at the $Y_C$ and $Y_N$ junctions, as illustrated in the inset. The quantized fractionalization of factor 1/3 suggests the generation of multiple wave packets (2$q$/3, 2$q$/9, …) toward the detector. This was experimentally confirmed as described in the main paper (see Fig. 3b). The red highlighted traces at $V_g$ = -0.3 V and $V_g$ = +0.26 V are shown as traces (i) and (ii), respectively, in Fig. 2b of the main paper.

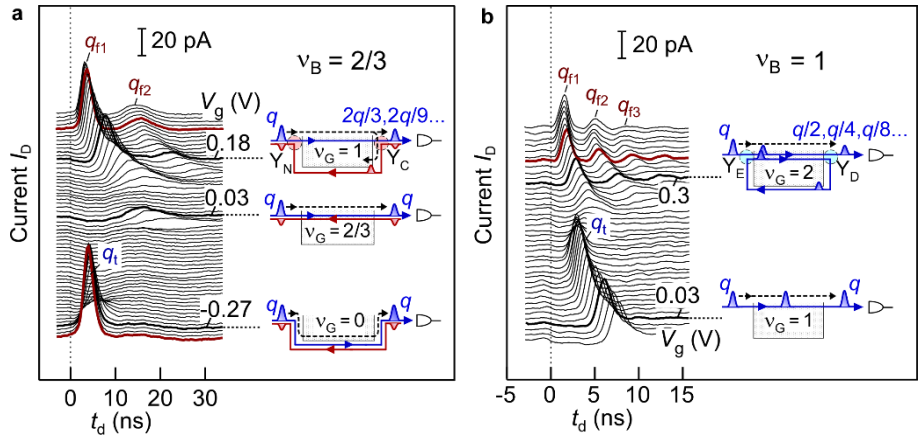

**Supplementary Figure 2. Charge waveforms in device #1. a**, Waterfall plot of current $I_D$ as a function of delay time $t_d$ for various $V_g$ values from -0.3 V (the bottom trace at $\nu_G = 0$) to +0.3 V (the top trace at $\nu_G \cong 1$) with step 0.01 V obtained at $B = 11.5$ T ($\nu_B = 2/3$). **b**, Current $I_D$ traces for various $V_g$ values from 0 V (the bottom trace at $\nu_G = 1$) to +0.4 V (the top trace at $\nu_G \cong 2$) with step 0.01 V obtained at $B = 7.5$ T ($\nu_B = 1$). The red highlighted traces are presented in Figs. 2b and 2c of the main paper.

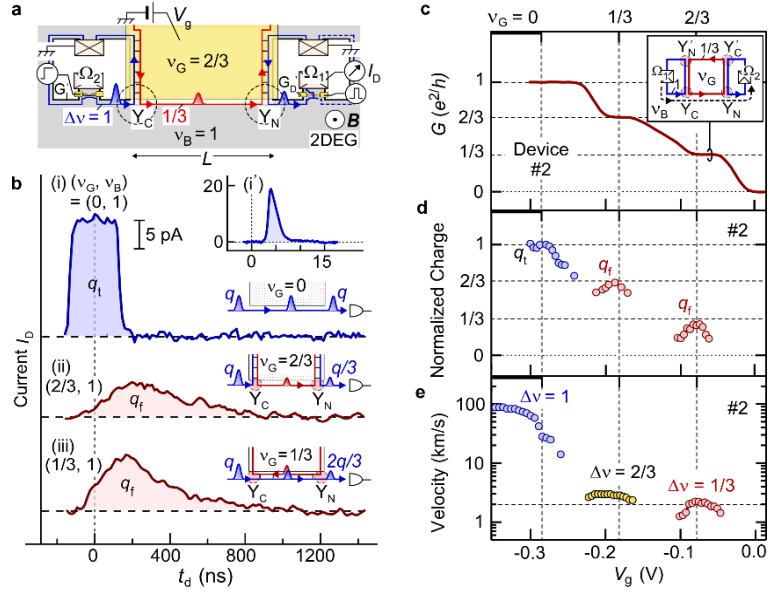

**Supplementary Figure 3. Fractionalization of charge wave packets. a**, Measurement setup for Corbino geometry in device #2. A wide QH region ($L$ = 420 μm) and junctions $Y_N$ and $Y_C$ were formed at $\nu_G$ = 2/3 and $\nu_B$ = 1. The initial wave packet was excited by applying a voltage step ($V_I$ = 15 mV) to gate $G_I$', and fractionalized wave packets were detected by applying a voltage pulse ($V_D$ = 20 mV) to gate $G_D$'. **b**, Typical charge waveforms obtained in current traces $I_D(t_d)$; reference trace (i) at $\nu_G$ = 0 ($V_g$ = -0.3 V), trace (ii) for fractionalized packet at $\nu_G$ = 2/3 ($V_g$ = -0.08 V), and trace (iii) at $\nu_G$ = 1/3 ($V_g$ = -0.18 V) under $\nu_B \cong$ 1 ($B$ = 9.5 T) and $t_w$ = 260 ns. The trace (i') in the inset shows a reference trace obtained with $t_w$ = 0.08 ns. Propagation of charge wave packets are illustrated in the insets. **c-e**, $V_g$-dependence of two-terminal conductance $G$ in **c**, charges $q_t$ and $q_f$ normalized by $q_t$ ($\cong$ 8000$e$) at $V_g$ = -0.3 V in **d**, and charge velocities for $\Delta\nu$ = 1/3, 2/3, and 1 channels in **e**. The corresponding $\nu_G$ is shown on the top scale. The inset to **c** shows charge fractionalizations under $\nu_G$ = 2/3 and $\nu_B$ = 1.

The data for quantized fractionalization in the integer QH regime are shown in Supplementary Fig. 2b with $I_D$ profiles measured in a wide range of $V_g$ from 0 V ($\nu_G \sim$ 1) to +0.4 V ($\nu_G \sim$ 2) at $\nu_B$ = 1 ($B$ = 7.5 T). A single wave packet was observed at approximately $\nu_G$ = 1, where a single $\Delta\nu$ = 1 channel was formed under the gate, as shown in the inset. By contrast, multiple peaks appeared at approximately $\nu_G$ = 2. This can be understood by the multiple fractionalizations between junctions $Y_E$ and $Y_D$, as illustrated in the inset. The red highlighted trace at $V_g$ = +0.34 V is shown in Fig. 2c of the main paper.

**Supplementary Note 3: Waveforms obtained from device #2**
As described in the main paper, an extremely slow propagation was observed for the $\Delta\nu$ = 1/3 interface channel with $\nu_G$ = 2/3 in the gated region. Because the wave packet was broadened significantly, it was difficult to identify multiple peaks in the measurement using device #1. Hence, we used the setup shown in Supplementary Fig. 3a, where $\nu_G$ = 2/3 and $\nu_B \cong$ 1 states were formed in the gated and bulk regions, respectively, at $B$ = 9.5 T. Charge $q$ generated with gate $G_I$' experiences fractionalizations first at junction $Y_C$ and then at $Y_N$ before reaching the detector gate $G_D$'. In this study, we focused on the fractionalized charge $q$/3 travelling in the interface channel $\Delta\nu$ = 1/3 (red line) formed between the $\nu_G$ = 2/3 and $\nu_B \cong$ 1 regions. The other charge 2$q$/3 fractionalized at $Y_C$ was absorbed in the grounded ohmic contact and therefore did not affect the measurement. This enables us to focus on the transport in the $\Delta\nu$ = 1/3 channel.

Trace (i) in Supplementary Fig. 3b shows the reference waveform obtained with $\nu_G$ = 0, where a single $\Delta\nu$ = 1 channel was formed between the injector and detector. Trace (ii) shows the wave packet obtained with $\nu_G$ = 2/3 for studying fractionalization, as illustrated in the inset. The fractionalized wave packet in trace (ii) is significantly delayed and broadened as compared with trace (ii) in Fig. 2b of the main paper. To obtain a reasonable signal-to-noise ratio, we set the width of the detector pulse, $t_w$ = 260 ns, which was comparable with the width of this fractionalized wave packet. This large $t_w$ was directly reflected in the width of the reference wave packet in (i); otherwise, a much narrower wave packet was observed for $t_w$ = 0.08 ns, as shown for trace (i') in the inset. We evaluated the reference charges $q_t$ and fractionalized charge $q_f$ with the same $t_w$, as shown in Supplementary Fig. 3d. The normalized charge $q_f/q_t$ approached 1/3 at $\nu_G$ = 2/3. Hence, the fractionalization ratio remains unchanged even when the wave packet is significantly delayed and distorted, as summarized in Fig. 4c of the main paper.

It should be noted that a clear wave packet was observed in trace (iii) of Supplementary Fig. 3b taken at $\nu_G$ = 1/3, where a different type of complex $\Delta\nu$ = 2/3 channel was formed between the $\nu_G$ = 1/3 and $\nu_B$ = 1 regions, as shown in the inset. The normalized charge for this packet was approximately 2/3, as shown in Supplementary Fig. 3d. Furthermore, the data set above supports fractionalization factor 1/3 and charge conservation in the system.

Moreover, the quantized fractionalization is consistent with the dc conductance measurement shown in Supplementary Fig. 3c. The two-terminal conductance between ohmic contacts $\Omega_1$ and $\Omega_2$ with other ohmic contacts floating is plotted as a function of $V_g$. The clear plateau of $G = e^2/3h$ at $\nu_G$ = 2/3 ($V_g$ = -0.08 V) indicates full equilibration in the complex $\Delta\nu$ = 2/3 channels. This can be understood by the 1/3 charge fractionalization at junction $Y_C$ in path $\Omega_1$ - $Y_C$ -$Y_N$ - $\Omega_2$, as shown in the inset.

Supplementary Fig. 3e shows the velocity estimated from the time-of-flight. Whereas the velocity of the edge channel $\Delta\nu$ = 1

between $v_G = 0$ and $v_B = 1$ was comparable to those in previous reports[38,39,43], the velocity of the $\Delta v = 1/3$ interface channel between the $v = 1$ and 2/3 regions was particularly slow at ~2 km/s (at $B = 9.5$ T), suggesting a large geometric capacitance between the channel and the gate, as discussed in the main paper.

**Acknowledgements**
The authors thank T. Hata and Y. Tokura for their beneficial discussions. This study was supported by the Grants-in-Aid for Scientific Research (JP15H05854, JP19H05603) and the Nanotechnology Platform Program of the Ministry of Education, Culture, Sports, Science and Technology, Japan.